\documentclass[11pt,a4paper]{article}
\usepackage[T1]{fontenc}
\usepackage[utf8]{inputenc}
\usepackage[english]{babel}
\usepackage{amsmath}
\usepackage{amssymb}
\usepackage{amsthm}
\usepackage{times}
\usepackage{a4}
\usepackage[hang,stable]{footmisc}

\usepackage{bbm}
\usepackage{color}
\usepackage{graphicx}
\usepackage{subfig}
\usepackage{caption}
\newcommand{\abs}[1]{\vert #1 \vert}
\newcommand{\D}{\mathrm{d}}

\begin{document}
\ifx\href\undefined\else\hypersetup{linktocpage=true}\fi

\title{Gravitationally induced inhibitions of dispersion 
       according to a modified Schr\"odinger-Newton equation
       for a homogeneous-sphere potential}
\author{Domenico Giulini and Andr\'e Gro{\ss}ardt       \\
        Center of Applied Space Technology and Microgravity\\
        University of Bremen             \\
        Am Fallturm 1                    \\
        D-28359 Bremen, Germany          \\
        and                              \\
        Institute for Theoretical Physics\\
        Leibniz University Hannover      \\
        Appelstrasse 2                   \\
        D-30167 Hannover, Germany}

\date{}

\maketitle

\begin{abstract}
\noindent
We modify the time dependent Schr\"odinger-Newton equation by
using a potential for a solid sphere suggested by J\"a\"askel\"ainen
(J\"a\"askel\"ainen 2012 \textit{Phys. Rev. A} \textbf{86} 052105)
as well as a hollow-sphere potential.
Compared to our recent paper
(Giulini and Gro{\ss}ardt 2011 \textit{Class. Quantum Grav.} \textbf{28} 195026)
where a single point-particle, i.\,e. a Coulomb potential, was considered
this has been suggested to be a more realistic model for a molecule.
Surprisingly, compared to our previous results, inhibitions of
dispersion of a Gaussian wave packet occur
at even smaller masses for the solid-sphere potential, given that
the width of the wave packet is not exceeded by the radius of the sphere.
\end{abstract}

%\begin{small}
%\setcounter{tocdepth}{3}
%\tableofcontents
%\end{small}
%\newpage

\section{Introduction}
In \cite{Giulini:2011} we investigated the time dependent 
Schr\"odinger-Newton equation (henceforth abbreviated 
SN~equation):
\begin{equation}
\label{eq:SchroedingerNewton}
\mathrm{i}\hbar\partial_t\Psi(t,\vec x)=
\left(
-\frac{\hbar^2}{2m}\Delta
+( \Phi \ast \vert\Psi\vert^2 )(t,\vec{x})
\right)
\Psi(t,\vec x)\,,
\end{equation}
with a potential term that is the spatial convolution
of the absolute-value squared of the wave function with
the Coulomb potential
$\Phi(r) = -G m^2 / r$:
\begin{equation}
\label{eqn:convolution}
( \Phi \ast \vert\Psi\vert^2 )(t,\vec{x}) =
-Gm^2\int\frac{\vert\Psi(t,\vec y)\vert^2}%
{\abs{\vec x-\vec y}}\,d^3y \,.
\end{equation}

Our numerical and
analytical consideration of the SN equation in \cite{Giulini:2011}
showed that inhibitions of dispersion due to the gravitational
self-interaction occur at mass values
of about $6.5 \times 10^9$\,u
for a given wave packet width of $0.5\,\mu\mathrm{m}$.
This result was contrary to a previous one by Salzman and Carlip 
\cite{Salzman:2006,Carlip:2008},
where inhibitions of dispersion were claimed to occur at mass values
more than six orders of magnitude smaller, namely at
$m\approx 1600\,\mathrm{u}$. Somewhat loosely,
this behaviour of a
wave packet initially shrinking in width was called a ``collapse''
by Salzman and Carlip and we adopt this nomenclature.

As we pointed out in the Summary of \cite{Giulini:2012}, more consideration
has to be given to the question concerning the application of the SN
equation to realistic systems, like molecules.
In a recent paper \cite{Jaeaeskelaeinen:2012} J\"a\"askel\"ainen suggested that
the gravitational self-interaction for a molecule should be modelled
by the potential of a solid sphere of radius $R$
\begin{equation}
\label{eqn:solid-sphere potential}
 \Phi(r) = \left\{\begin{array}{ll} -\frac{G m^2}{R} \left(\frac{3}{2}-\frac{r^2}{2 R^2}\right), & \mbox{if }r < R\\ -\frac{G m^2}{r}, & \mbox{if }r \geq R \;, \end{array}\right.
\end{equation}
which has to be put into equation (\ref{eq:SchroedingerNewton}) instead of the
Coulomb potential.

J\"a\"askel\"ainen points out that in our case, for a wave packet width of
$0.5\,\mu\mathrm{m}$, the results obtained for this potential deviate
from the results for the Coulomb potential for masses beyond $10^{10}$\,u.

We will next briefly review the rationale behind such a modification
and then turn to the quantitative changes it implies for our previous
analysis~\cite{Giulini:2011}. For comparison we will also consider the case
of a hollow sphere.

\section{How to separate the SN equation} \label{sec:separation}

In this section we briefly discuss how a one-particle SN equation 
for the centre of mass is obtained by separation from an $N$-particle 
SN equation, as claimed in~\cite{Jaeaeskelaeinen:2012}.

Consider the $N$-particle SN equation
\begin{align}
\mathrm{i} \hbar \dot{\Psi}_N(\vec{r}^N) &= 
\left[ - \sum_{i=1}^N \frac{\hbar^2}{2 m_i} \Delta_{\vec{r}_i} 
+ V_\text{EM}(\vec{r}^N) 
+ U_\text{G}[\Psi_N(\vec{r}^N)] \right] 
\Psi_N(\vec{r}^N), \label{eqn:N-particle-SN} \\
V_\text{EM}(\vec{r}^N) &= \sum_{i=1}^N \sum_{j=1}^{i-1} \frac{q_i q_j}{\abs{\vec{r}_i 
- \vec{r}_j}} \\
U_\text{G}[\Psi_N(\vec{r}^N)] &=
 -G \sum_{i=1}^N \sum_{j=1}^N m_i m_j \int \frac{\abs{\Psi_N(\vec{r'}^N)}^2}{\abs{\vec{r}_i 
- \vec{r_j'}}} \; \D {V'}^N \nonumber\\
&= -G \sum_{i=1}^N \sum_{j=1}^N m_i m_j \int \frac{P_j(\vec{r_j'})}{\abs{\vec{r}_i 
- \vec{r_j'}}} \; \D^3 r_j' \;. \label{eqn:UGrav}
\end{align}
Here $\vec{r}^N$ indicates the set of all $N$ coordinates $\vec{r}_i$, 
$m_i$ the mass of the $i$-th particle and $\Delta_{\vec{r}_i}$ is the 
Laplacian with respect to the $i$-th coordinate. By $P_i$ we denote 
the $i$-th marginal distribution that one obtains by integrating 
$\vert\Psi_N(\vec{r}^N)\vert^2$ over all $(N-1)$ factors $\mathbb{R}^3$ that 
parametrise the positions of all particles except the $i$-th.

The intuitive picture underlying the gravitational potential term 
$U_\text{G}$ is as follows: 
\emph{Each particle is under the influence of a Newtonian 
gravitational potential that is sourced by an active gravitational
mass-density to which each particle contributes proportional to its
probability density in position space as given by the marginal 
distribution of the total wave function:}
\begin{equation}
  \rho^\text{mat}(\vec{x}) = \sum_{j=1}^N m_j P_j(\vec{x})
= \sum_{j=1}^N m_j \int \abs{\Psi_N(\vec{r}^N)}^2 \delta(\vec{r_j'}-\vec{x}) \; \D {V'}^N\,.
\end{equation}

Although each particle lives in its own copy of $\mathbb{R}^3$ they all ``feel'' the same matter density $\rho^\text{mat}$. In particular, the 
self-interaction of each particle with its own field is included.

The gravitational interaction of all $N$ particles
is then given by
\begin{align}
 U_\text{G}[\Psi_N(\vec{r}^N)] 
&= -G\ \sum_{i=1}^N \int \frac{m_i \, \rho^\text{mat}(\vec{x})}{\abs{\vec{r}_i - \vec{x}}} \; \D^3 x \nonumber \\
&= -G\ \sum_{i=1}^N \sum_{j=1}^N \int \frac{m_i \, m_j \, P_j(\vec{x})}{\abs{\vec{r}_i - \vec{x}}} \; \D^3 x\,.
\end{align}

Note that this is the $N$-particle SN equation as it was stated by
Di\'{o}si~\cite{Diosi:1984}, which is different from that given by
J\"a\"askel\"ainen~\cite{Jaeaeskelaeinen:2012}. In~\cite{Jaeaeskelaeinen:2012}
the \emph{mutual} gravitational interactions are said to be negligible
compared to electromagnetic interactions and only the diagonal terms of
$U_\text{G}$ are considered. But in situations where the width 
of the individual marginal distributions is large compared to the 
mutual distances of their centres these contributions are of the same 
order of magnitude as the diagonal terms.

We emphasise that in this treatment there is a \emph{twofold} 
difference between the electromagnetic and the gravitational 
interaction of the particles. First, self-interactions of 
particles with their own \emph{electromagnetic} field are not   
considered. Second, whereas the electromagnetic interaction is 
local, the gravitational interaction is non-local in the sense 
explained above. And it is this non-locality that is responsible 
for a non-trivial contribution of the overall gravitational field 
to the dynamics of the centre of mass. 

We now introduce the centre of mass coordinates
\begin{equation}
 \vec{r} = \sum_{i=1}^N \frac{m_i \vec{r}_i}{M} \;, 
\quad M = \sum_{i=1}^N m_i \;, 
\quad \vec{\rho}_i = \vec{r}_i - \vec{r} \;\; (i=0,...,N-1)
\end{equation}
and make the separation ansatz
$\Psi_N(\vec{r}^N) = (m_N/M)^{3/2} \; \psi(\vec{r}) \chi(\vec{\rho}^{N-1})$,
where the prefactor comes from the integral measure and allows us to 
normalise all wave functions to one. Then $V_\text{EM}$ only depends on the 
relative coordinates $\vec{\rho}_i$.

Substituting the separation ansatz into the expression (\ref{eqn:UGrav}), 
$U_\text{G}$ can be written as
\begin{equation}
U_\text{G} 
\approx -G \sum_{i=1}^{N-1} \sum_{j=1}^{N-1} m_i m_j \int \frac{\abs{\psi(\vec{r'})}^2\;
 \abs{\chi(\vec{\rho'}^{N-1})}^2}{\abs{\vec{r} + \vec{\rho}_i - \vec{r'} 
- \vec{\rho_j'}}} \;\D^3 \rho_1' \cdots \D^3 \rho_{N-1}' \D^3 r' \;,
\end{equation}
where we assumed $N$ to be a large number, such that all 
terms involving the $N$-th particle can be neglected. 
They give only $(2N-1)$ out of $N^2$ contributions.%
\footnote{All contributions can be considered of the same order 
of magnitude. Otherwise just \emph{choose} the particle that 
yields the smallest contribution as the $N$-th particle.} 
Let
\begin{equation}
P^\text{rel}_i(\vec{\rho_i}) = \int  \abs{\chi(\vec{\rho}^{N-1})}^2 \;\D^3 
\rho_1 \cdots \D^3 \rho_{i-1} \D^3 \rho_{i+1} \cdots \D^3 \rho_{N-1}
\end{equation}
be the marginal distribution of the relative wave function $\chi$
for the $i$-th particle and rename the integration variable $\vec{\rho_j'}$ to
$\vec{\rho'}$. Then, in the above approximation, 
\begin{equation}
U_\text{G} 
= -G \int \abs{\psi(\vec{r'})}^2 \sum_{i=1}^{N-1} \sum_{j=1}^{N-1} m_i \int \frac{m_j\;
 P^\text{rel}_j(\vec{\rho'})}{\abs{\vec{r} + \vec{\rho}_i - \vec{r'} 
- \vec{\rho'}}} \;\D^3 \rho' \D^3 r' \,.
\end{equation}
Now note that $m_j P_j^\text{rel} = \rho_j^\text{mat}$ is simply the 
matter density of the $j$-th particle relative to the centre of mass.
We can therefore substitute the inner integral with the gravitational 
potential of the $j$-th particle, which is
\begin{equation}
\phi_j(\vec{x}) = -G M \; \int 
\frac{\rho_j^\text{mat}(\vec{y})}{\abs{\vec{x}-\vec{y}}} \; \D^3 y\,.
\end{equation}
We then obtain
\begin{equation}
U_\text{G} 
= \int \abs{\psi(\vec{r'})}^2 \; \sum_{i=1}^{N-1} \sum_{j=1}^{N-1} \frac{m_i}{M}
 \phi_j(\vec{r} + \vec{\rho}_i - \vec{r'}) \; \D^3 r' \,.
\end{equation}
If now we assume that the extent of the molecule is much smaller than the width 
of the wave function, such that the wave function does not change much over a 
distance $\abs{\vec{\rho}_i}$, i.\,e. 
$\abs{\psi(\vec{r'} - \vec{\rho}_i)}^2 \approx \abs{\psi(\vec{r'})}^2$, 
we can neglect the shift by $\vec{\rho}_i$. The sum over all $m_i$ then 
just yields the total mass $M$ and the one particle potentials
$\phi_j$ sum up to yield the full gravitational potential $\Phi$. We finally get
\begin{equation}
U_\text{G}[ \psi(\vec{r}) ]
= \int \abs{\psi(\vec{r'})}^2 \; \Phi(\vec{r} - \vec{r'}) \; \D^3 r' \,.
\end{equation}

The SN equation (\ref{eqn:N-particle-SN}) then separates into the ordinary 
electromagnetic multi-particle Schr\"odinger equation for the
relative coordinates, for which the solution is taken to be given by the
present lump of matter, and the SN equation (\ref{eq:SchroedingerNewton})
for the centre of mass wave function $\psi(\vec{r})$ with the modified
gravitational potential $\Phi$ for the matter at hand.

\section{Results}

\subsection{Solid-sphere potential}\label{sec:solid-sphere}

\newcommand{\pltw}{175pt}
\begin{figure}[t]
\subfloat[$m = 5 \times 10^9$ u]{%
\includegraphics[viewport=29 36 286 185,width=\pltw]{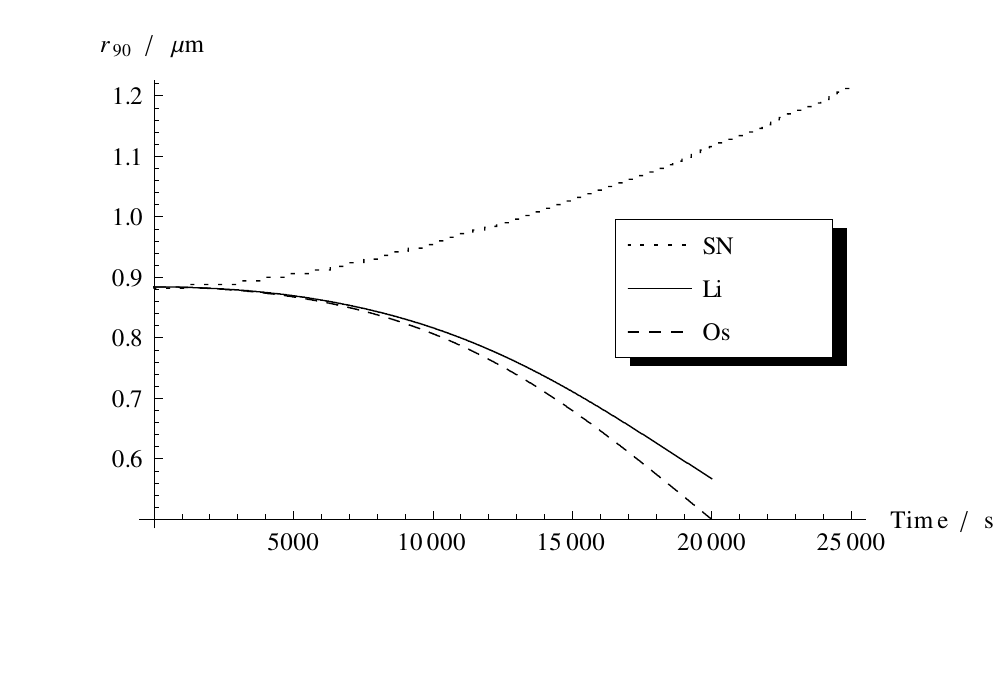}} \quad
\subfloat[$m = 7 \times 10^9$ u]{%
\includegraphics[viewport=29 35 286 177,width=\pltw]{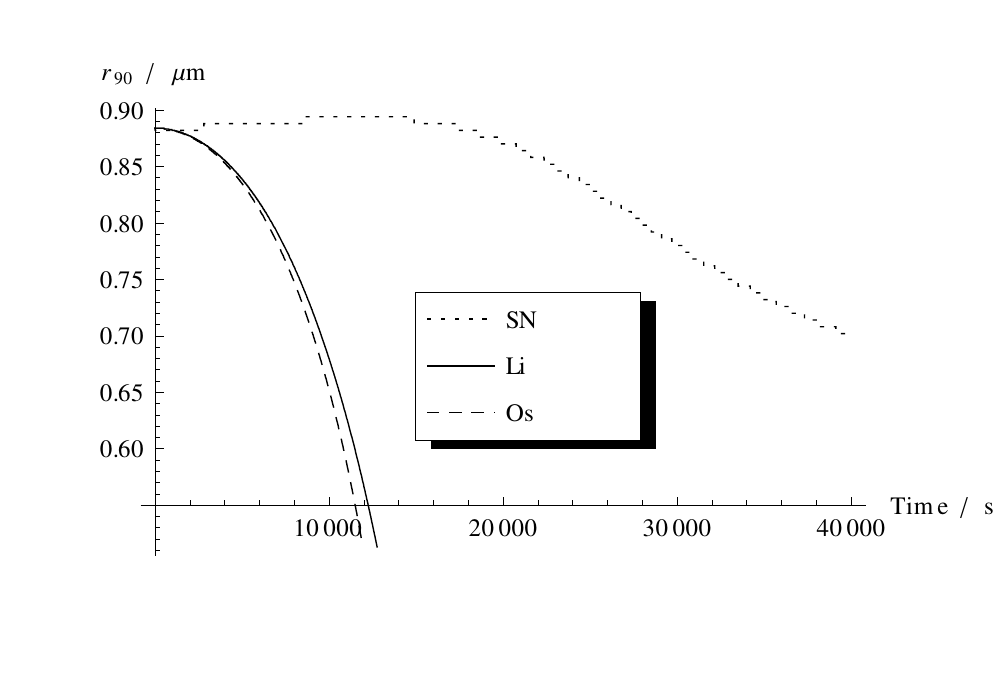}} \\
\subfloat[$m = 10^{10}$ u]{%
\includegraphics[viewport=29 36 286 187,width=\pltw]{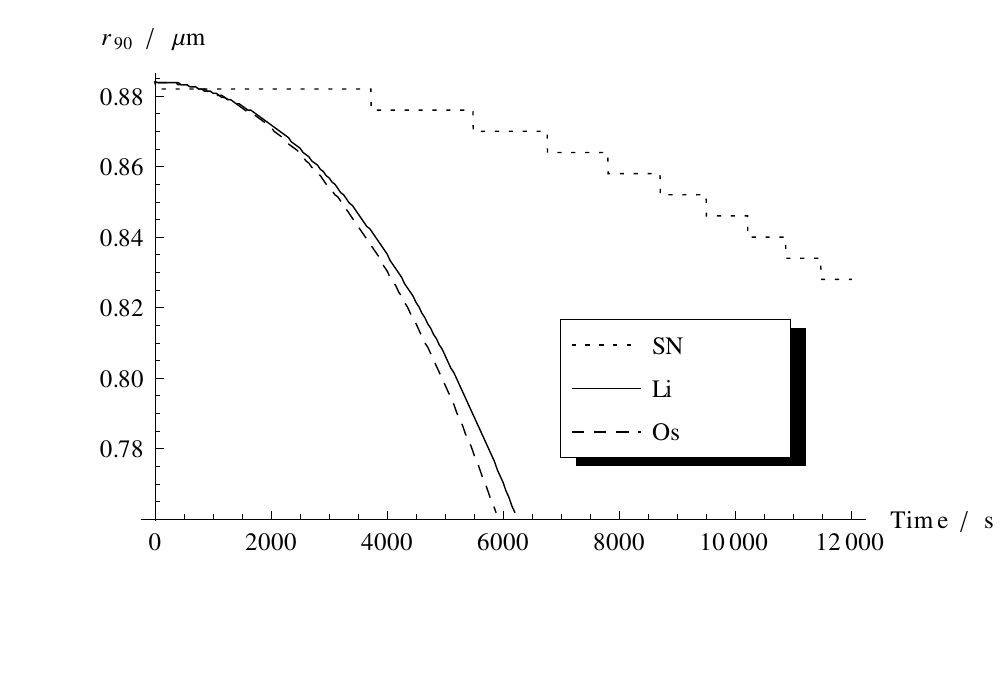}} \quad
\subfloat[$m = 10^{11}$ u]{%
\includegraphics[viewport=29 35 286 177,width=\pltw]{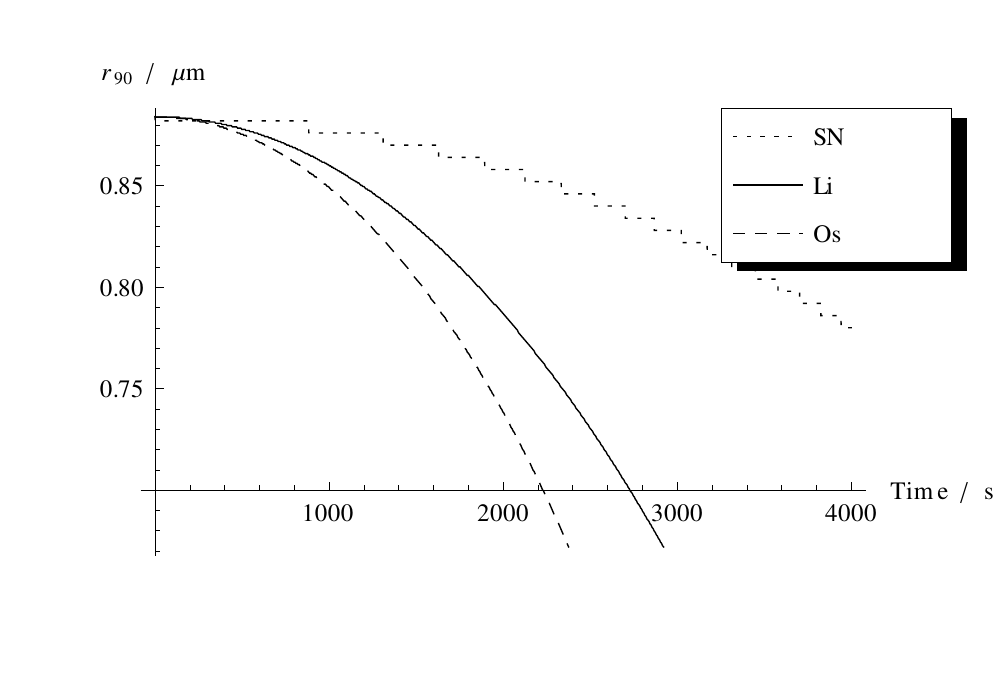}}
 \caption{This plot shows the radius $r_{90}$ within which 90\,\% of the
probability lie plotted against time for several masses for the SN equation with Coulomb potential (SN) and with a solid-sphere potential for both the density of lithium (Li) and osmium (Os).}
 \label{fig:plot_r90}
\end{figure}

\begin{figure}[t]
\centering
\subfloat[All radii]{%
\includegraphics[viewport=1 1 286 158,height=180pt]{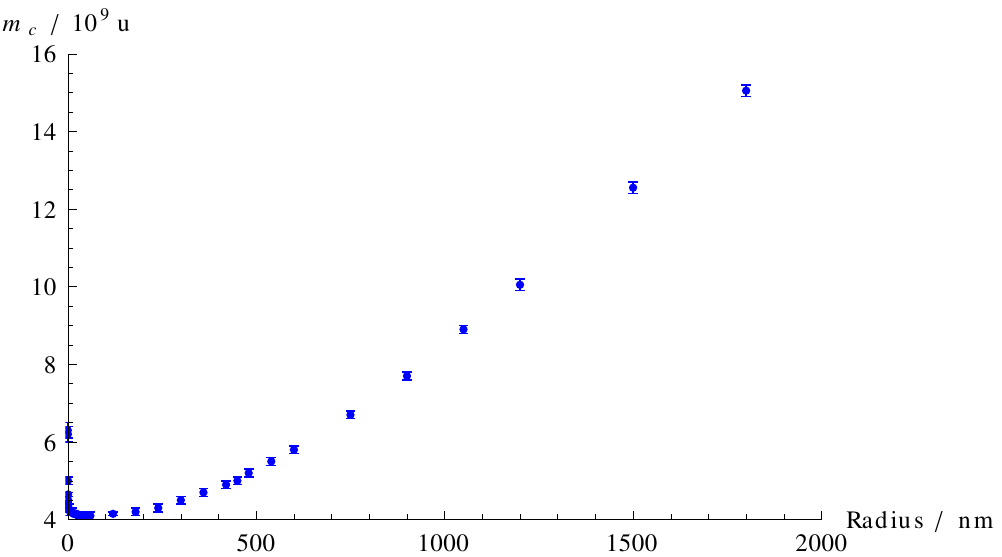}} \\
\subfloat[Zoom into region of small radii]{%
\includegraphics[viewport=1 1 286 158,height=180pt]{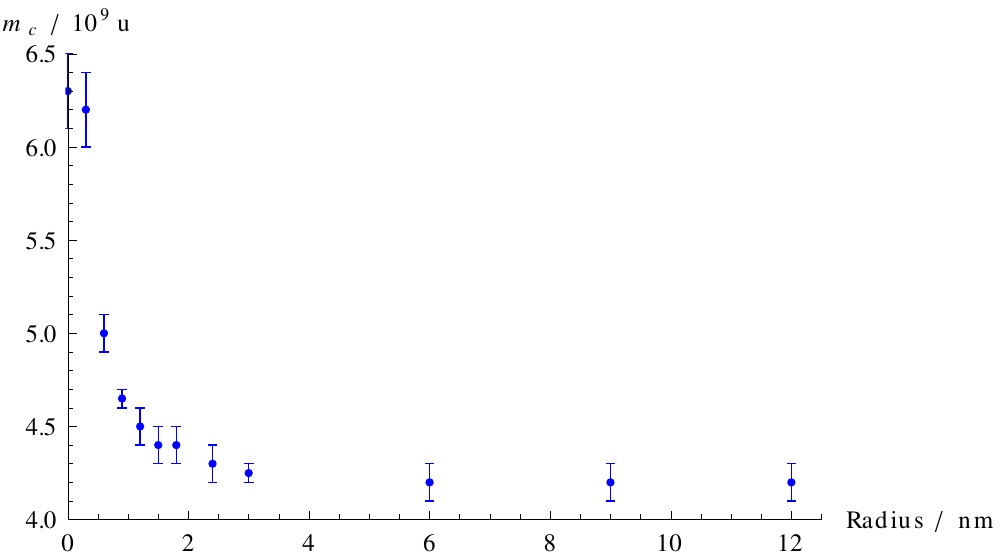}}
 \caption{The critical mass beyond which the wave packet reveals 
a shrinking behaviour is plotted against the radius of the 
solid sphere in the potential term.}
 \label{fig:plot_m_r}
\end{figure}

We study the evolution of the spherically symmetric SN equation making use of the same numerical
methods used in \cite{Giulini:2011}, simply replacing the potential by the
solid-sphere potential (\ref{eqn:solid-sphere potential}). All results refer
to an initial Gaussian wave packet
\begin{equation}
\label{eq:GaussPacket}
\Psi(r,t=0) = (\pi a^2)^{-3/4} \, \exp \left(-\frac{r^2}{2 a^2}\right)
\end{equation}
with a width of $a = 0.5\,\mu \mathrm{m}$.

First we look at the evolution of wave packets if the density of the solid
sphere corresponds to the density of either lithium (534\,kg\,$\mathrm{m}^{-3}$)
or osmium (22610\,kg\,$\mathrm{m}^{-3}$), cf. \cite{Jaeaeskelaeinen:2012}.
To illustrate the results, the radius within which 90\,\% of the probability
density are contained is plotted against time in figure \ref{fig:plot_r90}
for several masses. In addition to the results for the density of lithium (Li)
and osmium (Os) the results for the SN equation with Coulomb potential obtained
in \cite{Giulini:2011} are also plotted.

For all masses considered the modified potential results
in an even faster collapsing wave packet. The shrinking behaviour also sets
in for smaller masses, namely for $5 \times 10^9$\,u the modified solution
shrinks while the SN equation solution is still spreading for this mass.

To get a better impression of this behaviour, let us take a look at the effect
that the radius of the solid-sphere potential has on the mass for which the
collapsing
behaviour of the wave packet sets in. We thus ask the following question:
given a sphere of radius $R$, what mass $m_c$ do we have to distribute
homogeneously ``within'' this sphere\footnote{But note that the gravitational
interaction is always weighted with the probability density of the Gaussian
wave packet whose width
is always fixed to $0.5\,\mu \mathrm{m}$.} for a collapse of the wave packet
to take place?

In \cite{Giulini:2011} we found that for a $0.5\,\mu \mathrm{m}$ wave packet
a collapse is observed for masses of
about $6.5 \times 10^9$\,u and higher.
In figure \ref{fig:plot_m_r} this critical mass value is plotted depending on the
radius of the solid sphere.

For small radii of several nanometers this critical mass reduces to
about two-thirds of the value obtained in \cite{Giulini:2011} and it stays smaller
than this value until the size of the solid sphere exceeds the
width of the wave packet significantly.\footnote{As the separation in section
\ref{sec:separation} only holds for small radii compared to the extent of
the wave packet, the results for larger radii have to be regarded with some
scepticism, anyway.}
There seems to be an extremal value somewhere in between 20 and 100\,nm.

\subsection{Why does $R>0$ not lead to a diminishment of the collapse mass?}
\label{sec:total energy}

\begin{figure}[t]
\centering
\includegraphics[viewport=0 0 288 169,height=160pt]{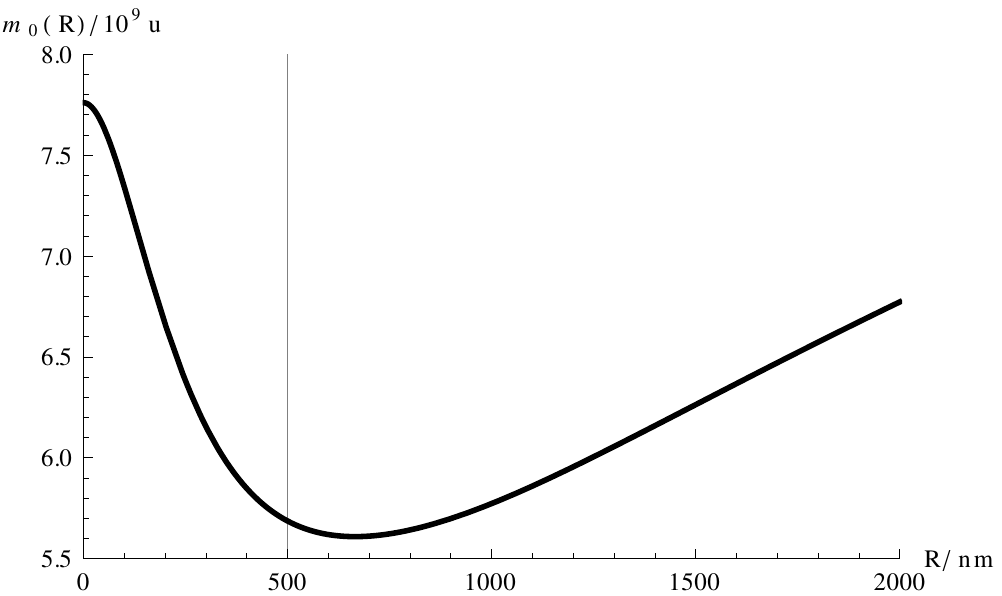}
 \caption{Plot of the mass $m_0(R)$ for which the total energy is zero.}
 \label{fig:m0 von R}
\end{figure}

The behaviour described in the previous paragraph seems peculiar, as naively one
would expect the gravitational field to be the weaker, and therefore the collapse mass
to be the higher, the larger the solid-sphere radius gets. But the ascertained
radius-dependence of the collapse mass becomes more comprehensible by an analysis
of the total energy value of the SN equation.

We performed a similar analysis in section 3.2 of \cite{Giulini:2011} where we
indicated the total energy for the SN wave packet:
\begin{equation}
\label{eqn:energy}
\mathcal{E} = T + \frac{1}{2} V
= \frac{\hbar^2}{2 m} \int \D^3 x \, \abs{\vec{\nabla} \psi(t,\vec{x})}^2
+ \frac{1}{2} \int \D^3 x \, \abs{\psi(t,\vec{x})}^2 \, U(t,\vec{x}) \, ,
\end{equation}
where $U(t,\vec{x}) = ( \Phi \ast \vert\Psi\vert^2 )(t,\vec{x})$.

Harrison \emph{et al}~\cite{Harrison:2003} pointed out that for the SN equation
a collapse can only be observed if the total energy is positive. The second
time derivative of the second moment $Q = \int \vec{x}^2 \abs{\psi}^2$ is given by
\begin{equation}
 \ddot{Q} = \frac{1}{m} \, \left( 4 T - 2 \int \D^3 x \, \abs{\psi(t,\vec{x})}^2 \,
 \vec{x} \cdot \vec{\nabla} U(t,\vec{x}) \right)
\end{equation}
for an arbitrary potential $U$. If the potential fulfils a poisson equation
$\Delta U \sim \abs{\psi}^2$ it takes the form
\begin{equation}
 \ddot{Q} = \frac{1}{m}\, (4 T + V) = \frac{1}{m} \, (4 \mathcal{E} - V),
\end{equation}
which is strictly positive for positive energy.

Although this relation does not hold for the modified potentials, the mass value for which
the total energy equals zero might nevertheless provide an insight of the qualitative
behaviour of the equation. We thus plot this mass value $m_0$ against the radius $R$ of
the solid-sphere in figure \ref{fig:m0 von R}. As one can see, this mass value indeed
decreases significantly within the width of the wave packet and starts to increase only
when the radius of the solid sphere exceeds this width.

Thus, qualitatively the behaviour of the zero-total-energy mass-value
coincides with the behaviour
of the collapse mass.

\subsection{Hollow-sphere potential}

At least for the experimental situation of interferometry with fullerene molecules,
it seems a more realistic model to consider a hollow-sphere potential
\begin{equation}
\label{eqn:hollow-sphere potential}
 \Phi(r) = \left\{\begin{array}{ll} -\frac{G m^2}{R}, & \mbox{if }r < R\\ 
-\frac{G m^2}{r}, & \mbox{if }r \geq R \end{array}\right.
\end{equation}
instead of the solid-sphere potential~(\ref{eqn:solid-sphere potential}).
Let us therefore repeat the analysis from section~\ref{sec:solid-sphere} for
the potential~(\ref{eqn:hollow-sphere potential}).

\begin{figure}[t]
\centering
\includegraphics[viewport=6 8 285 160,height=180pt]{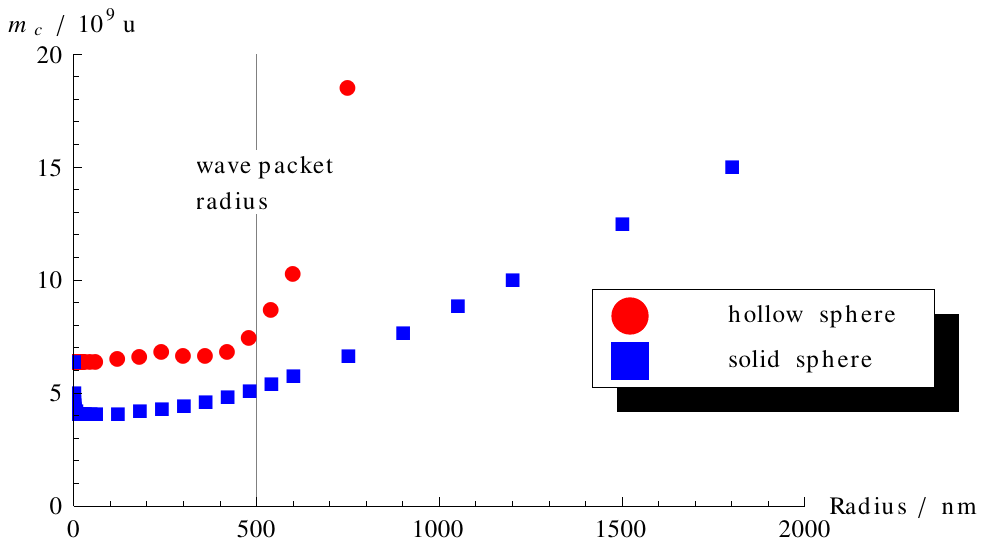}
 \caption{The critical mass beyond which the wave packet reveals a 
shrinking behaviour is plotted against the radius of the hollow (red dots) 
and solid (blue squares) sphere in the potential term.}
 \label{fig:plot_m_r_hohl}
\end{figure}

The result is plotted in figure~\ref{fig:plot_m_r_hohl}. As one can see, the
collapse mass does not decrease but remains almost constant as long as the radius
of the hollow sphere does not exceed the wave packet width. For larger radii
the collapse mass increases fast.

\begin{figure}[t]
\centering
\includegraphics[viewport=0 0 288 169,height=160pt]{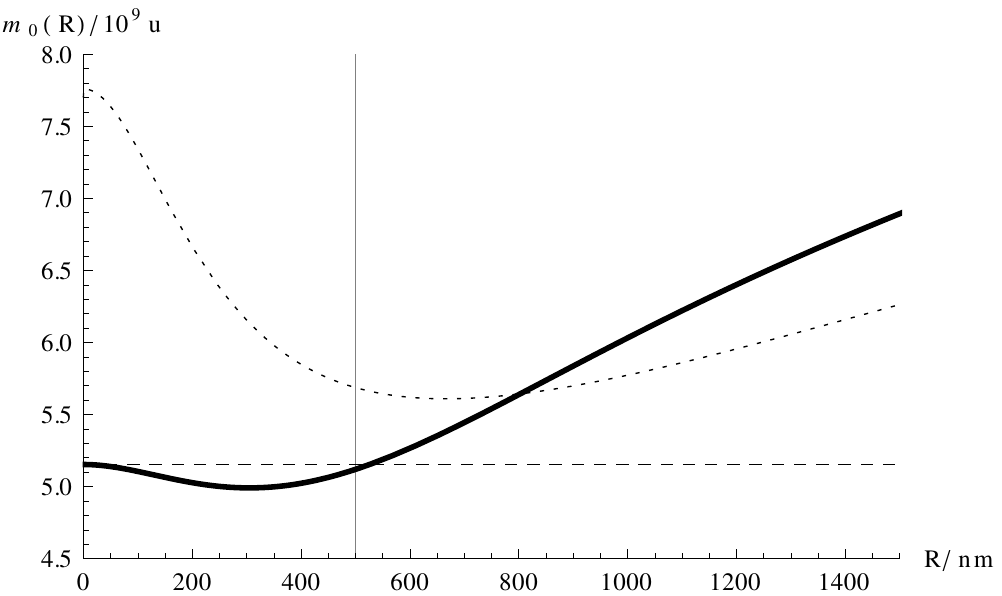}
\caption{Plot of the mass $m_0(R)$ for the Coulomb 
potential (dashed straight line), the solid sphere (dotted line) 
and the hollow sphere (solid line).}
 \label{fig:m0 von R 2}
\end{figure}

This can also be understood if, again, we look at the total energy
(see figure \ref{fig:m0 von R 2}).
Contrary to the solid sphere, $m_0$ for the hollow sphere converges
against the SN value for $R \to 0$
and compared to the solid-sphere value (dotted line) it only decreases very weakly
before increasing for radius values above the width of the wave packet.

\section{Conclusion}

If one does not take the simpler point of view we took in \cite{Giulini:2011}
and models large molecules by a point-particle SN equation, and instead
assumes the modified equation given in \cite{Jaeaeskelaeinen:2012}
to provide a more realistic description for such molecules, our results confirm that
there are indeed deviations between both equations, as J\"a\"askel\"ainen suggested.

However, whereas one naively might expect that a smaller matter density leads to a weaker
gravitational interaction and therefore to a behaviour that is closer to the
free Schr\"odinger equation, our results show that the converse
is true. As long as the radius of the solid-sphere potential does not significantly
exceed the extent of the wave packet, the effect of inhibitions of dispersion
is even increased compared to the SN equation.
The analysis of the total energy given in section \ref{sec:total energy}
provides an insight why this somewhat peculiar behaviour shows up.

Such an increase of the effect cannot be observed for the hollow-sphere potential,
but there is also no significant decrease, as long as the radius of the hollow
sphere does not exceed the extent of the wave packet.

The question why there should be a self-gravitational interaction while there
is no electromagnetic self-interaction term present in the Schr\"odinger equation
remains an issue that should be addressed (cf. the discussion in \cite{Giulini:2012}).
This issue arises even more distinctly
in the analysis given in section \ref{sec:separation} of this paper, where the
gravitational interactions are treated differently 
to the electromagnetic ones.

We conclude that taking into account the finite extent of the source
in the way suggested in~\cite{Jaeaeskelaeinen:2012} does not lead to
an attenuation of the effect of inhibitions of dispersion as discussed
in~\cite{Giulini:2011}.
The prospects for an experimental verification of self-gravitation of
quantum systems in future molecular
interferometry experiments remain unchanged.

\section*{Acknowledgements}
We gratefully acknowledge funding and support through the Centre for
Quantum Engineering and Space-Time Research ({\small{QUEST}})
at the Leibniz University Hannover.

%%%%%%%%%%%%%%%%%%%%%%%%%%%%%%%%%%%%%%%%%%%%%%%%%%%%%%%%%%%%%%%%%%%%%%

\bibliographystyle{iopart-num}
\footnotesize
%\bibliography{giulini_grossardt_sn_spherical}

\begin{thebibliography}{1}
\expandafter\ifx\csname url\endcsname\relax
  \def\url#1{{\tt #1}}\fi
\expandafter\ifx\csname urlprefix\endcsname\relax\def\urlprefix{URL }\fi
\providecommand{\eprint}[2][]{\url{#2}}

\bibitem{Giulini:2011}
Giulini D and Gro{\ss}ardt A 2011 {\em Class. Quant. Grav.\/} {\bf 28} 195026

\bibitem{Salzman:2006}
Salzman P~J and Carlip S 2006 A possible experimental test of quantized gravity
  arXiv:gr-qc/0606120 based on the Ph.D. thesis of Salzman: ``Investigation of
  the Time Dependent Schr\"odinger-Newton Equation'', Univ. of California at
  Davis, 2005

\bibitem{Carlip:2008}
Carlip S 2008 {\em Class. Quant. Grav.\/} {\bf 25} 154010

\bibitem{Giulini:2012}
Giulini D and Gro{\ss}ardt A 2012 {\em Class. Quant. Grav.\/} {\bf 29} 215010

\bibitem{Jaeaeskelaeinen:2012}
J\"a\"askel\"ainen M 2012 {\em Phys. Rev. A\/} {\bf 86} 052105

\bibitem{Diosi:1984}
Di{\'o}si L 1984 {\em Phys. Lett. A\/} {\bf 105} 199

\bibitem{Harrison:2003}
Harrison R, Moroz I and Tod K~P 2003 {\em Nonlinearity\/} {\bf 16} 101

\end{thebibliography}
\providecommand{\newblock}{}

\end{document}